\documentclass{amsart}
\usepackage{graphicx}
\newtheorem{lem}{Lemma}
\newtheorem{op}{Open Problem}
\newtheorem{thm}{Theorem}

\newtheorem{cor}{Corollary}
\newcommand{\pf}{{\noindent\bf Proof:\,}}

\begin{document}
\title{The maximum number of minimal codewords in long codes}
\author{A. Alahmadi}
\address{MECAA, Math Dept of King Abdulaziz University, Jeddah, Saudi Arabia }
\email{adelnife2@yahoo.com}
\author{R.E.L. Aldred}
\address{Department of Mathematics and Statistics,
University of Otago,
P. O. Box 56, Dunedin, New Zealand}
\email{raldred@maths.otago.ac.nz}
\author{R. dela Cruz}
\address{Division of Mathematical Sciences, SPMS, Nanyang Technological University, Singapore
\\and\\Institute of Mathematics, University of the Philippines Diliman, Quezon City, Philippines}
\email{ROMA0001@e.ntu.edu.sg}
\author{P. Sol\'e}
\address{Telecom ParisTech, 46 rue Barrault, 75634
Paris Cedex 13, France.\\and\\MECAA, Math Dept of King Abdulaziz University, Jeddah, Saudi Arabia }
\email{sole@enst.fr}
\author{C. Thomassen}
\address{Department of Mathematics,
Technical University of Denmark,
DK-2800 Lyngby, Denmark\\and\\MECAA, Math Dept of King Abdulaziz University, Jeddah, Saudi Arabia }
\email{C.Thomassen@mat.dt.u.dk}
%\date{} % Activate to display a given date or no date (if empty),
         % otherwise the current date is printed

\begin{abstract}
Upper bounds on the maximum number of minimal codewords in a binary code follow from the theory of matroids.
Random coding provide lower bounds. In this paper we compare these bounds with analogous bounds for the cycle code of graphs. This problem (in the graphic case) was considered in 1981 by Entringer and Slater who asked if a connected graph with $p$ vertices and $q$ edges can have only slightly more that $2^{q-p}$ cycles. The bounds in this note answer this in the affirmative for all graphs except possibly some that have fewer than $2p+3\log_2(3p)$ edges. We also conclude that an Eulerian (even) graph has at most $2^{q-p}$ cycles unless the graph is a subdivision of a $4$-regular graph that is the edge-disjoint union of two Hamiltonian cycles, in which case it may have as many as $2^{q-p}+p$ cycles.

\end{abstract}

\maketitle

\section{Introduction}
The codewords of a binary code form a poset under support inclusion. The codewords that are minimal for that partial order are called \emph{ minimal.}
They include the minimum weight codewords, but do not coincide, in general, with them. They occurred in decoding studies \cite{A}, and independently in secret sharing schemes based on codes \cite{AB}.
What is the number $M(C)$ of minimal codewords a binary linear code $C$ might have?
If $C$ has dimension $k$, an immediate upper bound, which we call the \emph{trivial upper bound}, is $M(C)\le 2^k-1.$ This bound is met with equality for {\em intersecting codes}, i.e. codes any pair of codewords of which intersect nontrivially \cite{CL}. Conversely, any code meeting that bound with equality is intersecting.

If $G$ is a connected graph on $p$ vertices with $q$ edges, then its cycle code $C(G)$ has length
$n=q,$ and dimension $k=q-p+1.$ The minimal codewords of $C(G)$ are exactly the incidence vectors of cycles (that is, circuits in the cycle matroid) in $G.$ Thus the above question restricted to the graphic case asks how many cycles a graph with $p$ vertices and $q$ edges can have. (We allow graphs to have multiple edges but no loops.) This question was raised in 1981 by Entringer and Slater \cite{ES} who observed that a graph with $p$ vertices and $q$ edges cannot have more than $2^{q-p+1}$ cycles because of the trivial upper bound. They gave examples showing that it may have slightly more that $2^{q-p}$ cycles and asked if these examples were close to best possible. In this paper we verify this for all graphs except possibly some that are sparse.

We view this in a coding theoretic context as follows: We denote by $\mu(R)$ the asymptotic exponent of the maximum of
$M(C)$ for families  of codes $C$
of rate $R,$ where $R$ is the  limsup of $k/n.$ Formally, let ${\bf  C}[n,k]$ denote
the set of all $[n,k]$ codes, and let
$$M(n,k)=\max \{M(C) :\, C \in {\bf C}[n,k]\}.$$

We now introduce

$$\mu(R)=\limsup_{n \rightarrow \infty} \frac{1}{n} \log_2 M(n,\lceil Rn \rceil) . $$

If, in this definition, we replace ${\bf  C}[n,k]$ by the cycle codes of graphs with $p$ vertices and $q$ edges (where $n=q,$ and $k=q-p+1$) we obtain $M_g(n,k), \mu_g(R)$, respectively.

 By the trivial upper bound, $\mu(R)\le R$. For $R\in [0,0.5]$ random coding shows that the bound is tight. For $R>0.5$ the trivial upper bound can be improved using matroid theory for $R\ge R_0$ with $R_0\approx 0.77.$

 In this paper we prove an upper bound for the number of cycles in a graph with $p$
 vertices and $q$ edges. That upper bound implies that

 $$\mu_g(R)=R$$

 for $R \leq 0.5$, and

$$\mu_g(R)=-(1-R)\log_2(1-R)$$

for $0.5 <R <1$.

 It also implies that every graph with more than $2p+O(\log(p))$ edges has fewer than $2^{q-p}$ cycles. The graph $2C_p$ (the cycle of length $p$ with every edge doubled) is $4$-regular, and has therefore $2p$ edges, and has precisely
$2^{q-p}+p= 2^p+p$ cycles. This shows that we cannot omit the $O(\log(p))$ term above. But we may be able to omit it if we the raise the bound $2^{q-p}$ slightly. We show that for this, it suffices to investigate a very special class of graphs, namely those $4$-regular graphs which are the union of two Hamiltonian cycles.

%%%%%%%%%%%%%%%%%%%%%
\section{Known bounds on $M(n,k), M(C)$ and $\mu(R)$}

In this section we review some known bounds.

\vspace{5mm}

If $C$ is an $[n,k]$ code, then we have $M(C)\le 2^k-1.$ We call this bound the \emph{trivial upper bound.}
It is easy to see that a binary linear code $C$ meets the the trivial bound with equality if and only if it is intersecting.
For, if $C$ is not intersecting, then two of its codewords, say $c$ and $d$ have disjoint supports.
Their sum $c+d$ is nonzero and non-minimal. Hence  $M(C)< 2^k-1.$
Conversely, if $C$ has a non-minimal codeword, then it can be written as a sum of at least two
disjoint support minimal codewords. Therefore $C$ is not intersecting.

\vspace{5mm}
The trivial upper bound implies that $\mu(R)\le R.$

\vspace{5mm}

It is proved in \cite{DSL} that, for a binary matroid on $n$
points of rank $n-k$ represented by an $[n,k ]$ code $C$ say, we have
$$M(C)\le { n \choose k-1} .$$

This implies that $\mu(R)\le H(R)$, where $H$ is Shannon's binary entropy function defined for $x\in [0,1]$ by
$$H(x)=-x\log_2 x -(1-x)\log_2 (1-x).$$
That upper bound is better than the trivial upper bound for $R> R_0\approx 0.77$ where $H(R_0)=R_0.$

\vspace{7mm}

By averaging arguments (random coding) it was shown in \cite[Cor. 2.5]{AB} that for $R<1/2,$ we have
$$\mu(R)\ge R,$$
and for $R > 1/2,$ we have
$$\mu(R)\ge H(R)-1+R.$$

\vspace{5mm}
%%%%%%%%%%%%%%%%%%%%%%%
 Finally in this section we state, as Theorem 1, a bound for $M(n,k)$  from  \cite[Theorem 5]{A}.
 We refer to this bound as the \emph{Agrell upper bound}.

{\thm For $\frac{k-1}{n} > \frac{1}{2}$ we have $$ M(n,k)  \le \frac{2^k}{4n(\frac{k-1}{n}-\frac{1}{2})^2}.$$ } \qed

\section{Cycle codes of graphs.}

It is a long-standing and difficult problem in graph theory to find the maximum number of
cycles a connected graph on $p$ vertices and with $q$ edges can have. This problem was raised by
Entringer and Slater \cite{ES} who observed that no connected graph $G$ can have more than $2^{q-p+1}$ cycles. This follows from the trivial upper bound because there is a
binary $[q, q-p+1]$ code $C(G)$ called the \emph{cycle
code} of the graph. Its codewords are defined on the edge set and are the indicator vectors of the
edge disjoint unions of cycles. The minimal codewords of  $C(G)$ are the indicator vectors of
the cycles of $G$. Entringer and Slater \cite{ES} also observed that there are graphs having slightly more than $2^{q-p}$ cycles and asked if this is (essentially) the maximum. Finally, Entringer and Slater \cite{ES} observed that the maximum is attained for cubic graphs.

The first bound significantly below the trivial bound was obtained by Aldred and Thomassen \cite{AT}
who proved that no connected graph $G$ can have more than $\frac{15}{16} 2^{q-p+1}$ cycles.
This is the best known upper bound for cubic graphs. But, for graphs of average degree $>4$ there
are better bounds, and in fact, the question by Entringer and Slater has been answered for all graphs
of average degree slightly greater than $4$. The Agrell bound (Theorem 2) immediately implies the following.
{\cor  If $G$ is a connected graph on $p$ vertices and with $q$ edges satisfying $q>2p$, then its number of cycles is at most
$$ \frac{q2^{q-p+1}}{(q-2p)^2}.$$  \qed}

We shall prove the following.

{\thm  If $G$ is a connected graph on $p$ vertices and with $q$ edges, and we write
$$q-1=(p-1)m+r$$

where $m,r$ are nonnegative integers, and
$0 \leq r <p-1,$

then its number of cycles is at most
$$ qm^{p-1-r}(m+1)^r. $$  \qed}

If we fix the ratio $q/p$ and let $p$ tend to infinity, then the bounds in Corollary 1 and Theorem 2
are essentially exponential functions. The exponential function in Theorem 2 is in a sense best possible, as we point out below. Also, both results answer the question of Entringer and Slater (asking if a graph can have significantly more than $2^{q-p}$ cycles) for graphs of average degree slightly more than $4$. Corollary 1 shows that counterexamples (if any) can have
at most $2p+O(\sqrt p)$ edges. Theorem 2 goes further and says that they must have at most $2p+O(\log(p))$ edges. It would be interesting to answer the question for all graphs of average degree at least 4.

\vspace{5mm}

Theorem 2 is a consequence of Theorem 3 below.

\vspace{7mm}

A {\bf path} is a graph with vertices $v_1,v_2, \ldots ,v_p$ and edges $v_1v_2, v_2v_3, \ldots ,v_{p-1}v_p.$
A {\bf multipath} is obtained from a path by replacing some edges with multiple edges. Let $f(q,p)$ denote the maximal number of paths from $v_1$ to $v_p$ in a multipath with $p$ vertices and $q$ edges. This maximum is attained if no two edge multiplicities differ by more than $1$. So, if $p-1$ divides $q$, then

$$f(q,p)=(q/(p-1))^{p-1}.$$

%\vspace{7mm}
{\lem If $x,y$ are vertices in a graph $G$ with $p$ vertices and $q$ edges, then $G$ has at most $f(q,p)$ paths from $x$ to $y$.}\\ \\

\pf The proof is by induction on $p$. If $p=2$ the statement is trivial. So assume $p>2$.
Let $d$ denote the degree of $x$, and let $x_1,x_2, \ldots ,x_d$ be the neighbors of $x$. (Some of these neighbors may be identical.) By induction, $G-x$ has at most $f(q-d,p-1)$ paths from $x_i$ to $y$, for each $i=1,2, \ldots ,d$. So $G$ has at most $df(q-d,p-1)$ paths from $x$ to $y$. As $f(q-d,p-1)$ is the number of paths between the ends in a multipath with $p-1$ vertices and $q-d$ edges, we may interprete $df(q-d,p-1)$ as the number of paths between the ends in a multipath with $p$ vertices and $q$ edges, where the first edge multiplicity in the multipath is $d$. By the maximum property of $f(q,p)$, we have

$$df(q-d,p-1) \leq f(q,p),$$

which completes the proof. \qed

\vspace{7mm}
{\thm Let $p,q$ be natural numbers $\geq 2$.

There exists a graph with $p$ vertices, $q$ edges, and at least $f(q-1,p)$ cycles.

If $G$ is any graph with $p$ vertices and $q$ edges, then $G$ has at most $$qf(q-1,p)$$ cycles.}
\vspace{7mm}

\pf

 By the definition of $f$, there exists a multipath with $p$ vertices, $q-1$ edges, and precisely $f(q-1,p)$ paths between the ends. If we add an edge between the ends we get a graph with $p$ vertices, $q$ edges, and at least $f(q-1,p)$ cycles.

To prove the last statement, consider any edge $e=xy$ in $G$. The number of cycles in $G$ containing $e$ is the number of paths in $G-e$ from $x$ to $y$. By Lemma 1, this number is at most $$f(q-1,p).$$

This completes the proof.\qed

\vspace{7mm}

Because of the logarithm in the definition of $\mu$, Theorem 3 gives the right $\mu$-value for the graphs of a fixed average degree. It is still interesting, though, to decide if the bound $qf(q-1,p)$ can be lowered to about $(q/p)^p$. Does the maximum number of cycles occur in graphs similar to $tC_p$? Are there graphs of average degree $2t$ without multiple edges that have the same number or a larger number of cycles?
These questions are open even for cubic graphs. As mentioned earlier, Entringer and Slater \cite{ES} observed that a cubic graph on $p$ vertices may have as many as
$2^{p/2}$ cycles. It has been open for several years if this is close to the right number. For planar cubic graphs this was verified by Aldred and Thomassen \cite{AT}. For general cubic graphs they lowered the trivial upper bound $2^{p/2+1}$ to $(15/16) \cdot 2^{p/2+1}$.

\vspace{3mm}

We can now answer the question by Entringer and Slater \cite{ES} for all graphs with average degree slightly more than $4$.

\vspace{3mm}

{\thm Let $G$ be a graph with $p$ vertices and $q$ edges. If $q>2p+3\log_2(3p)$, then $G$ has at most $2^{q-p}$ cycles.}

\vspace{3mm}

\pf
Consider first the case where $q=2p$. The upper bound on the number of cycles provided by Theorem 2 is $(2p) \cdot 3 \cdot 2^{p-2}$, which is more than $2^{q-p}=2^p$. However, if we increase $q$ to $2p+r$, then the upper bound increases to $(2p+r) \cdot 3 \cdot 2^{p-2} \cdot (3/2)^r $. This number is $\leq 2^{p+r}$ for $r>3\log_2(3p)$.

This completes the proof.\qed

\section{Applications to $\mu_g(R)$.}

We can now determine $\mu_g(R)$ completely.

\vspace{7mm}

{\thm For $0<R\leq 0.5,$ $$\mu_g(R)=R.$$

If $R$ is of the form $1-1/t$ where $t$ is a natural number $\geq 2$, then $$\mu_g(R)=-(1-R)\log_2(1-R).$$}

The function $\mu_g(R)$ is continuous and linear in each closed interval from $1-1/t$ to $1-1/(t+1)$, where $t$ is a natural number $\geq 2$.

\vspace{7mm}

\pf

Assume first that $0<R<0.5$. For any two natural numbers $p,r$ we let $C_{p,r}$ be obtained from $2C_p$ by subdividing one edge $r$ times. Then $C_{p,r}$ has $p+r$ vertices and $n=2p+r$ edges.
Thus the dimension of the cycle code is $k=p+1$, and the rate of the cycle code is $k/n=(p+1)/(2p+r)$. For each natural number $p$ we let $r$ be the largest natural number such that $k/n=(p+1)/(2p+r) \geq R$. Then $k=\lceil Rn \rceil$. Also, $r= \lfloor(p+1)/R-2p \rfloor$. Recall that the number of cycles in $2C_p$ and hence also in $C_{p,r}$ is $>2^p$. Substituting these values in the definition of $\mu_g$ and letting $p$ tend to infinity, we conclude that $\mu_g(R) \geq R$.

The trivial upper bound shows that this inequality is, in fact, an equality.
\vspace{7mm}

Consider next the case where $R=1-1/t$ for where $t$ is natural number $\geq 2$.
Let $tC_p$ be the cycle of length $p$ where each edge has been
duplicated $t$ times. This graph has $p$ vertices and $q=pt$ edges. The number of cycles in this graph is

$$t^p+p{t \choose 2},$$
the first term counting cycles of length $p$ and the second cycles of length $2.$
The graph is regular of degree $2t$. Hence it has $n=pt$ edges and has rate $(pt-p+1)/pt=1-1/t-1/pt$. Hence its cycle code is an $[n,Rn+1]$-code. If we delete an edge, then we get an $[n-1,\lceil R(n-1) \rceil]$-code. Deleting an edge reduces the number of cycles only slightly. Letting $p$ and hence also $n$ tend to infinity, we conclude that

$$\mu_g(R) \geq -(1-R)\log_2(1-R).$$

\vspace{7mm}

Now assume that $1-1/t <R < 1-1/(t+1)$, where $t$ is a (fixed) natural number $\geq 2$. then we let $G(p,R,r)$ denote the graph obtained from $tC_p$ by adding $r$ edges between neighboring vertices such that all edge multiplicities are $t$ or $t+1$. The resulting graph $G(p,R,r)$ has $p$ vertices and $n=pt+r$ edges.
Thus the dimension of the cycle code is $k=p(t-1)+r+1$, and the rate of the cycle code is $k/n=(p(t-1)+r+1)/(pt+r)$.

We first choose any $p$ so large that the rate of $tC_p$, which is $(pt-p+1)/pt=1-1/t-1/pt$, is smaller
than $R$. Then we let $r$ be the largest natural number such that the rate of $G(p,R,r)$, which is $(p(t-1)+r+1)/(pt+r)$, is smaller than or equal to $R$. That is, $r=-pt+\lfloor (p-1)/(1-R) \rfloor$. Then the cycle code of $G(p,R,r)$ is an $[n,\lceil Rn \rceil]$-code, where $n=pt+r=\lfloor (p-1)/(1-R) \rfloor$ is the number of edges of $G(p,R,r)$.

The number of cycles in $G(p,R,r)$ is $> t^{p-r}(t+1)^r$.

If we substitute these values in the definition of $\mu_g$ and let $p$ tend to infinity, then we conclude that

$$\mu_g(R) \geq ((1-R)(1+t)-1)\log(t)+(-t(1-R)+1)\log(t+1).$$

The right hand side is clearly a linear function. For $R$ equal to $1-1/t$ or $1-1/(t+1)$ the right hand side has the same values as the lower bounds we obtained for those two values of $R$. So we have obtained a lower bound for $\mu_g(R)$ which is continuous and piecewise linear.

We claim that this lower bound is also an upper bound. We used the graph $G(p,R,r)$ above. If we put $q=n=pt+r$, then $G(p,R,r)$ has the maximum number of cycles among those graphs with $q$ edges which are obtained from a cycle of length $p$ by duplicating edges. The graphs used to give the lower bound $f(q-1,p)$ in Theorem 3 are also graphs of this type. Hence $G(p,R,r)$ has at least $f(q-1,p)$ cycles. On the other hand, Theorem 3 says that any graph with $p$ vertices and $q$ edges has at most $qf(q-1,p)$ cycles. So no graph with $p$ vertices and $q$ edges has more than $q$ times as many cycles as $G(p,R,r)$. Hence the lower bound for $\mu_g(R)$ obtained from the graphs $G(p,R,r)$ is also an upper bound. \qed

\vspace{7mm}

The function $\mu_g(R)$ is less than the matroid upper bound for all $R$ and also less than the random upper bound for $R>0.5$. The Agrell upper bound gives the same upper bound on $\mu$ as the trivial upper bound. Figure 1 shows these bounds.

\section{Cycle codes of $4$-regular graphs}

Cycle codes of cubic (that is, $3$-regular) graphs have enjoyed particular attention because in order to answer the afore-mentioned question by Entringer and Slater \cite{ES}, it suffices to consider cubic graphs. The investigations in this paper indicate that the $4$-regular graphs also deserve attention. First of all, their cycle codes have rate $R \approx 0.5$, and this is the smallest value of $R$ for which the function $\mu_g(R)$ changes shape. Here we shall provide another reason. Although it is merely an observation we call it a theorem because of the striking exceptions that appear in the statement. A graph is \emph{Eulerian} if it is connected and all vertices have even degree. These graphs are particularly interesting in the present context because the vector $\bf{1}$ consisting of ones is a code word. Recall that a \emph{Hamiltonian cycle} in a graph is a cycle containing all vertices.

{\thm
If $G$ is an Eulerian graph with $p$ vertices and $q$ edges, then $G$ has at most $2^{q-p}$ cycles unless $G$ is a subdivision of a $4$-regular graph which is a union of two Hamiltonian cycles.}

\vspace{7mm}

\pf The map sending a codeword $\bf{x}$ into the codeword $\bf{x}+\bf{1}$ is a map from the cycle code into itself.
If the codeword of every cycle is mapped into a codeword which does not correspond to a cycle, then at most half of the $2^{q-p+1}$ codewords correspond to cycles, and the result follows. So we may assume that $G$ has a cycle $C$ such that both its codeword $\bf{x}$ and also the codeword $\bf{x}+\bf{1}$ correspond to cycles, say $C,C'$. Hence $G$ cannot have a vertex of degree $6$ or more. Each of $C,C'$ contains precisely two edges incident with each vertex of degree $4$. And each edge of $G$ belongs to precisely one of $C,C'$. So $C,C'$ is a partition of $G$ into two Hamiltonian cycles after the vertices of degree $2$ have been replaced by edges. \qed

\vspace{7mm}

\begin{figure}[h]
\begin{center}
\includegraphics[width=0.8\linewidth]{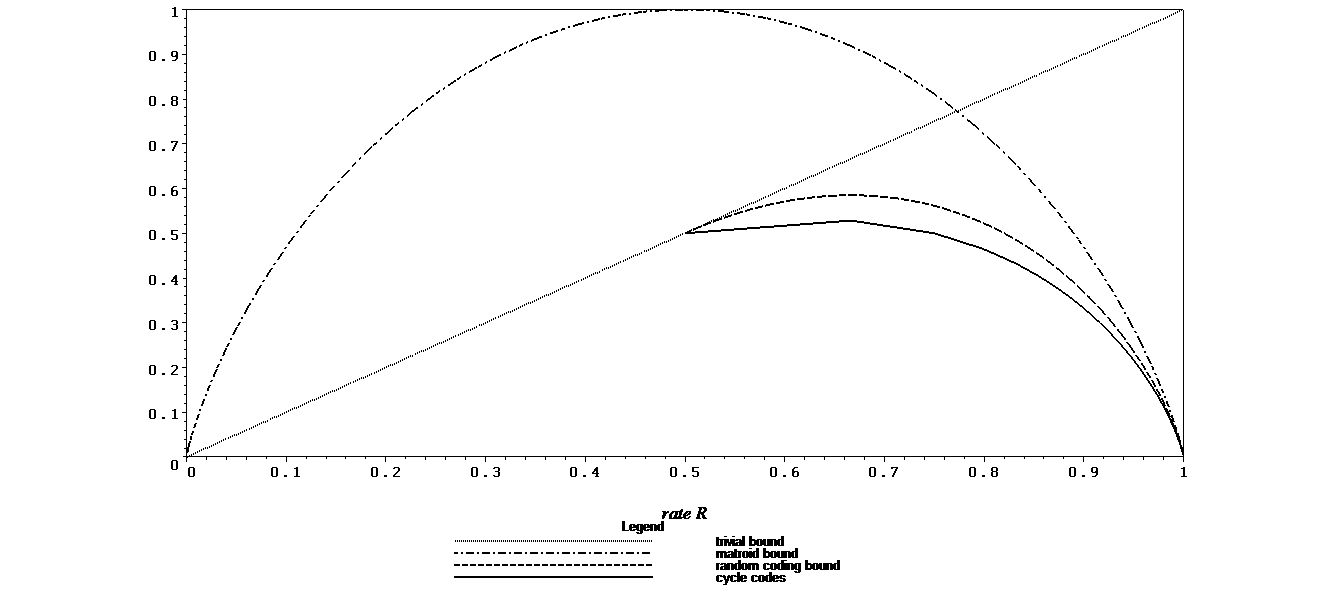}
\caption{$\mu(R)$ versus $R$}
\label{Fig. 1}
\end{center}
\end{figure}

\vspace{7mm}

\section{Conclusion and Open problems}

Figure 1 shows as a function of $R,$
\begin{itemize}
\item the trivial upper bound $R,$
\item the matroid upper bound $H(R)=-R\log_2(R)-(1-R)\log_2(1-R)$
\item the random lower bound $H(R)-1+R$
\item the  function $\mu_g(R)$ for cycle codes of graphs.
\end{itemize}

The function $\mu_g(R)$ is known exactly. It is $\mu_g(R)=R$ for $R\le 0.5,$ and $\mu_g(R)=-(1-R)\log_2(1-R)$ for $R> 0.5.$ The function $\mu$ seems more problematic.
 Thus $\mu(R)=R$ for $R\le 0.5,$ while for $R>0.5$ we only have the bounds
$$H(R)-1+R \le \mu(R)\le \min(R,H(R))$$ the best known upper bound for $R>0.5$.
These bounds can be generalized to linear codes over a non-binary alphabet.
%For a linear code meeting the trivial upper bound on $M(n,k)$ is equivalent to be intersecting. From \cite[Th. 2.3]{CZ} we know that the trivial upper bound is tight {\em constructively} for $R\le 0.156.$

By \cite{CL} long  linear intersecting codes can exist only for $R<0.283.$ Thus the codes of rate between that value and $0.5$ provided by random coding are ``almost'' intersecting.

\vspace{7mm}

While we have found $\mu_g(R)$ it is a wide open problem to find $\mu(R)$. Even the following seems nontrivial.

{\op Is $\mu(R)$ a continuous function of $R?$ Is it concave? For which $R$ is it maximum?}

\vspace{3mm}

Analogous questions have been considered by Manin \cite{M}.

\vspace{3mm}

We have answered Problem 1 for $\mu_g(R)$. This function has maximum for $R=2/3$. The $6$-regular graphs have rate $\approx 2/3$, so maybe also the $6$-regular graphs are worth studying in more detail.

\vspace{7mm}

As mentioned earlier, the $4$-regular graph $2C_p$ has $2^p+p$ cycles, and no $4$-regular graph has more than $2^{p+1}$ cycles, by the trivial upper bound.

{\op Does there exist a $4$-regular graphs with $p$ vertices and more than $2^p+p$ cycles? }

\vspace{3mm}
In case the answer is negative, then it seems that the $4$-regular graphs are the only regular graphs for which there is a simple expression for the maximum number of cycles.
\vspace{3mm}
{\op Does there exist a real number $c<2$ such that every $4$-regular graphs with $p$ vertices and no multiple edges has less than $c^{p+1}$ cycles? }

\vspace{3mm}

As mentioned above, $tC_p$ has $p$ vertices, $q=pt$ edges, and $t^p+p{t \choose 2}$ cycles.

 {\op Does there exist a graph with $p$ vertices, $q=pt$ edges (where $t$ is a natural number), and more than
 $t^p+p{t \choose 2}$ cycles?}

\vspace{3mm}

{\bf Acknowledgement:} The authors are indebted to G. D. Cohen and H. Randriam for helpful discussions.

\end{document}